\documentclass[conference]{IEEEtran}
\usepackage{epsfig}
\usepackage{amssymb}
\usepackage[font=footnotesize]{caption} 
\usepackage{latexsym,graphicx}
\usepackage{longtable}
\usepackage{algpseudocode} 
\usepackage{algorithm}
\usepackage{array}
\usepackage{mathrsfs}
\usepackage{color}
\usepackage{amsmath,graphicx}
\usepackage{relsize}
\usepackage{epsfig}
\usepackage{setspace}
\usepackage{cite}
\usepackage{hyperref}
\usepackage{amssymb}
\usepackage{caption}
\usepackage{subcaption}
\usepackage[normalem]{ulem}
\usepackage{soul,xcolor}
\def\BibTeX{{\rm B\kern-.05em{\sc i\kern-.025em b}\kern-.08em
    T\kern-.1667em\lower.7ex\hbox{E}\kern-.125emX}}

 \makeatletter
 \let\NAT@parse\undefined
 \makeatother

\begin{document}
%
\title{Square Deviation Based Symbol-Level Selection for Virtual Full-Duplex Relaying Networks} 
%
%
\author{$\text{Jiancao Hou}^{\dagger}$, $\text{Sandeep Narayanan}^{\dagger}$, $\text{Na Yi}^{\star}$, $\text{Yi Ma}^{\star}$, and $\text{Mohammad Shikh-Bahaei}^{\dagger}$
\\{\small $^{\dagger}$Centre for Telecommunications Research, King's College London, London, UK}
\\{\small $^{\star}$Institute for Communication Systems, University of Surrey, Guildford, UK}
\\{\small e-mail:\{jiancao.hou, sandeep.kadanveedu, m.sbahaei\}@kcl.ac.uk, \{n.yi, y.ma\}@surrey.ac.uk}


}
\markboth{} {Shell \MakeLowercase{\textit{et al.}}: Bare Demo of
IEEEtran.cls for Journals}\maketitle

\begin{abstract}
In this paper, a symbol-level selective transmission for virtual full-duplex (FD) relaying networks is proposed, which aims to mitigate error propagation effects and improve system spectral efficiency. The idea is to allow two half-duplex relays, mimicked as FD relaying, to alternatively serve as transmitter and receiver to forward the source's messages. In this case, each relay predicts the correctly decoded symbols of its received frame, based on the generalized square deviation method, and discard the erroneously decoded symbols, resulting in fewer errors being forwarded to the destination. Then, a modified maximum \textit{a posteriori} receiver at the destination is provided to eliminate the inter-frame interference and identify the positions of discarded symbols from the relays. In addition, the diversity-multiplexing trade-off (DMT) for our proposed scheme is also analysed. It is found that our proposed scheme outperforms the conventional selective decode-and-forward (S-DF) relaying schemes, such as cyclic redundancy check based S-DF and threshold based S-DF, in terms of DMT. Moreover, the bit-error-rate performance are also simulated to confirm the DMT results. 
\end{abstract}


\IEEEpeerreviewmaketitle
 

\section{Introduction}
Cooperative communication has attracted much attention in recently years due to its capability of serving as a virtual multi-antenna based system to combat fading and improve spectral efficiency. Moreover, it can also be utilized to extend the communication coverage. In the literature, there are many relaying protocols being considered, and one of their practical ones is decode-and-forward (DF) relaying, where the relay node decodes the received signal sent from the source node and forwards the regenerated one to the destination node\cite{Laneman2004,Chen2006,Lai2006}. It is shown that this protocol simplifies the power control and allow for reprocessing of the decoded signal. However, the erroneously decoded bits or symbols may be forwarded to the destination node, which results in the error propagation effects and degrades the system performance. 

On the other hand, conventional relaying normally works in half-duplex (HD) constrained scenarios to avoid the co-channel interference. However, this consideration may implicitly limit the communication capacity, where every data frame needs to be transmitted via two successive time slots or two different frequency bands. To efficiently utilize the limited radio resources, many relaying protocols have been proposed in the literature, and one of their notable examples is named virtual full-duplex (FD) relaying, or also called successive relaying, in \cite{Oechtering2004,Ribeiro2004}. This protocol mimics FD relaying \cite {Ju2009,Duarte2010,Naslcheraghi2017} and allows the source node continuously transmit the information data for every channel use, while two HD relay nodes alternatively serve as transmitter and receiver to forward the source's messages. In this case, diversity-multiplexing trade-off (DMT) can be used to indicate the system performance, and the optimal performance can be achieved if the two HD relay can perfectly decode the source's messages \cite{Fan2007}. However, this assumption may be hard to realize due to strong inter-relay interference and/or inter-cell interference.

In order to combat with the effect of erroneously decoding, the authors in \cite{Kobravi2007,Wang2009,Bobarshad2009,Shikh-Bahaei2004,Tian2011,Shadmand2010,Hu2012} presented the adaptive retransmission request schemes to guarantee the perfectly decoding at the relay nodes at the cost of time delay and signalling overhead \cite{Shadmand2009,Zarringhalam2009,Olfat2005,Shojaeifard2011}. Alternatively, the authors in \cite{Sun2011,Basar2014} proposed the space-time coding schemes to assist the relay nodes to guarantee the decoding quality, where the decoding error can still happen especially for higher modulations. Without considering the adaptive retransmission request or the space-time coding with lower modulations, the authors in \cite{Wicaksana2011} analysed the DMT performance with the combination of frame-level selective decode-and-forward (S-DF) relaying protocol. In this case, the relay nodes directly decode the source node's messages and treat the inter-relay interference as noise, and if there exist decoding errors, they will stop forwarding the received frames in the next time slot. Such scheme can be likened to the conventional S-DF relaying with cyclic redundancy check (CRC). Based on the analysis in \cite{Wicaksana2011}, the DMT performance is significantly degraded if the relay nodes cannot perfectly decode source's messages. This is because that a few erroneously decoded symbols stop the relay nodes forwarding the correctly decoded symbols to enjoy spatial diversity gain. This motivates us to propose the symbol-level selective transmission method to improve the system performances.

In detail, instead of discarding the whole transmission frame after CRC fails as in \cite{Wicaksana2011}, the relay nodes in our proposed scheme predict the correctly decoded symbols per frame based on square deviation principle \cite{Kay1993} and then forward them to the destination node, resulting in fewer errors being propagated. In the literature, log-likelihood ratio (LLR) based symbol-level selective transmission has been proposed in \cite{Al-Habian2011,Kwon2010}, where the correctly decoded symbols are selected by comparing their LLR values with a pre-determined threshold. To precisely select the qualified symbols, appropriate LLR threshold needs to be obtained. However, such appropriate LLR threshold is hard to find \cite{Yi2015}. In addition, the relay nodes with the selection methods in \cite{Al-Habian2011,Kwon2010} need to inform the destination node the positions of discarded symbols. This will cost extra signalling overhead. In contrast, our proposed symbol-level selection method simplify the decision threshold design, and with the proposed modified maximum \textit{a posteriori} (MAP) detector, the positions of discarded symbols can be estimated at the destination node. Apart from these, the DMT performance of our proposed scheme is also analysed in this paper. By comparing with the DMT of existing S-DF based scheme, our proposed scheme can offer higher diversity gain while make multiplexing gain stay the same. Furthermore, computer simulations in terms of bit-error-rate (BER) are also implemented to confirm the DMT results.

 

\section{System Model}
\begin{figure}[t] 
\begin{center}
\epsfig{figure=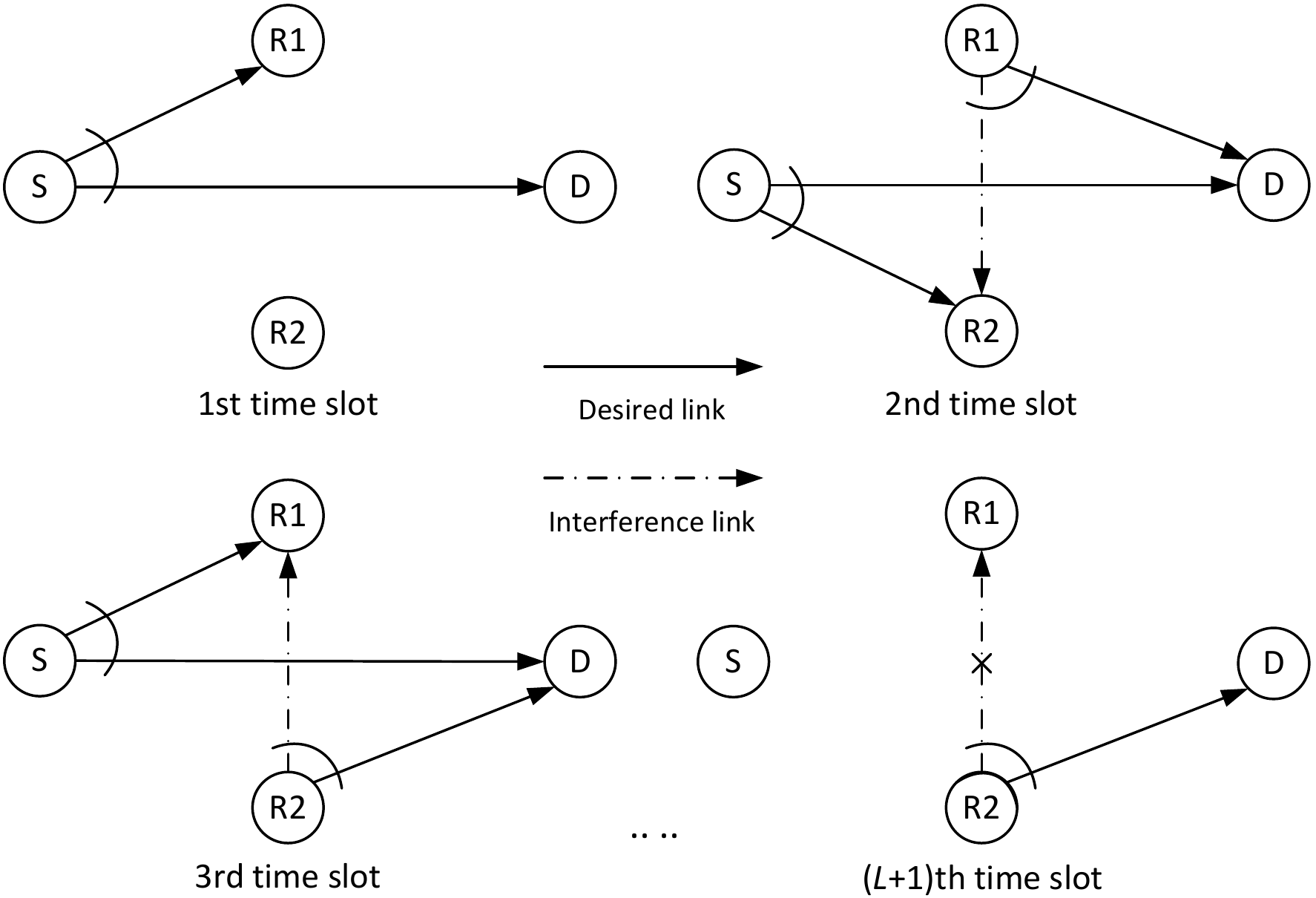,scale=0.45,angle=0}
\end{center}
\caption{Illustration of virtual FD relaying network.}\label{F1}
\end{figure}

We assume discrete-time block fading channels, which remain static over each transmission time slot. Consider a virtual FD relaying network with one source node (S), two HD relay nodes (R1/R2) and one destination node (D) as illustrated in Fig.~\ref{F1}, 
where each node only has a single antenna. The source node encodes the information bits $\mathbf{b}$ using a turbo-like encoder, e.g., a 1/2 rate serial concatenated convolutional code, to generate the coded bits $\mathbf{c}$. Then, $\mathbf{c}$ are mapped onto $\mathbf{x}$ transmission symbols based on $Q$-ary modulation scheme. Subsequently, $\mathbf{x}$ is divided into $L$ frames, and without loss of generality $L$ is assumed to be even. The source node broadcasts $L$ data frames in $L$ time slots, and two relay nodes alternatively transmit and receive the data sent from the source node until the $(L+1)^{\mathrm{th}}$ time slot. Specifically, for the even time slots, relay one sends its decoded frame to relay two and the destination node, meanwhile, the source node sends the new frame to relay two and the destination node; for the odd time slots, relay two sends its decoded frame to relay one and the destination node, meanwhile, the source node sends the new frame to relay one and the destination node.  

Let $\mathbf{y}_{\mathrm{R}j}(l)\in\mathcal{C}^{M}$ be the received signal vector in the $l^{\mathrm{th}}$ time slot at the $j^{\mathrm{th}}, j\in\{1,2\}$, relay node, and $\mathbf{y}_{\mathrm{D}}(l)\in\mathcal{C}^{M}$ be the received signal vector in the $l^{\mathrm{th}}$ time slot at the destination node, where $M$ is the number of symbols per frame. Then, we have
\begin{equation}\label{eq01}
\mathbf{y}_{\mathrm{R}j}(l)=\sqrt{P}h_{\mathrm{S},\mathrm{R}j}(l)\mathbf{x}_{\mathrm{S}}(l)+\sqrt{P}h_{\mathrm{R}}(l)\mathbf{x}_{\mathrm{R}\overline{j}}(l-1)+\mathbf{v}_{\mathrm{R}j}(l),
\end{equation}
\begin{equation}\label{eq02}
\mathbf{y}_{\mathrm{D}}(l)=\sqrt{P}h_{\mathrm{S},\mathrm{D}}(l)\mathbf{x}_{\mathrm{S}}(l)+\sqrt{P}h_{\mathrm{R}\overline{j},\mathrm{D}}(l)\mathbf{x}_{\mathrm{R}\overline{j}}(l-1)+\mathbf{v}_{\mathrm{D}}(l),
\end{equation}
where $h_{\mathrm{S},\mathrm{R}j}(l)$, $h_{\mathrm{R}}(l)$, $h_{\mathrm{S},\mathrm{D}}(l)$, and $h_{\mathrm{R}\overline{j},\mathrm{D}}(l)$ are the complex channel coefficients in the $l^{\mathrm{th}}$ time slot for S-R link, R-R link, S-D link, and R-D link, respectively, and $\overline{j}$ is complementary of $j$, e.g., if $\mathrm{R}j$ represents R1, $\mathrm{R}\overline{j}$ will be R2, vise versa; $\mathbf{x}_{\mathrm{S}}(l)\in\mathcal{C}^{M}$ is the $l^{\mathrm{th}}$ frame transmitted in the $l^{\mathrm{th}}$ time slot at the source node; $\mathbf{x}_{\mathrm{R}\overline{j}}(l-1)\in\mathcal{C}^{M}$ is the $(l-1)^{\mathrm{th}}$ frame transmitted in the $l^{\mathrm{th}}$ time slot at the $\overline{j}^{\mathrm{th}}$ relay node; $P$ is the transmit power for each transmission node; $\mathbf{v}_{\mathrm{R}j}(l)\in\mathcal{C}^{M}$ and $\mathbf{v}_{\mathrm{D}}(l)\in\mathcal{C}^{M}$ are the additive white Gaussian noise (AWGN) with zero mean and covariance of $\mathbf{I}$ for the $j^{\mathrm{th}}$ relay and destination nodes, respectively, where $\mathbf{I}$ is the $M\times M$ identity matrix.   

From the channel model described above, if the relay nodes can perfectly decode the messages sent from the source node, the MISO equivalent form for the entire $L+1$ time slots transmission from the source to the destination with the help of the relays can be expressed as
\begin{equation}\label{eq03}
\mathbf{Y}=\sqrt{P}\mathbf{H}\mathbf{X}+\mathbf{V},
\end{equation}
where $\mathbf{Y}\in\mathcal{C}^{(L+1)\times M}$ is the received signal matrix at the destination node; $\mathbf{X}=[\mathbf{x}_\mathrm{S}(1),\mathbf{x}_\mathrm{S}(2),\ldots,\mathbf{x}_\mathrm{S}(L)]^{T}$ is the $L\times M$ transmitted frames from the source node, and $(\cdot)^{T}$ is the transpose of a matrix; $\mathbf{V}\in\mathcal{C}^{(L+1)\times M}$ is AWGN matrix; $\mathbf{H}\in\mathcal{C}^{(L+1)\times L}$ is given by
\begin{IEEEeqnarray}{ll}\label{eq04}
\mathbf{H}=\nonumber\\
\left[\begin{array}{ccccc}
h_{\mathrm{S,D}}(1) & 0 & \cdots & 0 & 0 \\
h_{\mathrm{R1,D}}(2) & h_{\mathrm{S,D}}(2) & \cdots & 0 & 0 \\
0 & h_{\mathrm{R2,D}}(3) & \cdots & 0 & 0 \\
\vdots & \vdots & \ddots & \vdots & \vdots \\
0 & 0 & \cdots & h_{\mathrm{R1,D}}(L) & h_{\mathrm{S,D}}(L) \\
0 & 0 & \cdots & 0 & h_{\mathrm{R2,D}}(L+1) \end{array} \right].\nonumber\\
\end{IEEEeqnarray}



\section{The Proposed Symbol-Level Selective Transmission Scheme}
Our proposed scheme relies on the square deviation method (see \cite{Kay1993}) to predict the position of correctly decoded symbols in a frame. This leads to a low-complexity and high-efficiency symbol-level selection method at the relay nodes. To elaborate, we first transfer \eqref{eq01} to a symbol-wise equation as
\begin{equation}\label{eq09}
y^{(m)}_{\mathrm{R}j}(l)=\sqrt{P}h_{\mathrm{S},\mathrm{R}j}(l)x^{(m)}_{\mathrm{S}}(l)+z^{(m)}_{\mathrm{R}j}(l),~\forall m,
\end{equation}
where $y^{(m)}_{\mathrm{R}j}(l)$ and $x^{(m)}_{\mathrm{S}}(l)$ are the $m^{\mathrm{th}}$ elements in $\mathbf{y}_{\mathrm{R}j}(l)$ and $\mathbf{x}_{\mathrm{S}}(l)$, respectively; $z^{(m)}_{\mathrm{R}j}(l)\triangleq{\sqrt{P}h_{\mathrm{R}}(l)x^{(m)}_{\mathrm{R}\overline{j}}(l-1)+v^{(m)}_{\mathrm{R}j}(l)}$ is the inter-relay interference plus AWGN, where $x^{(m)}_{\mathrm{R}\overline{j}}(l-1)$ and $v^{(m)}_{\mathrm{R}j}(l)$ are the $m^{\mathrm{th}}$ elements in $\mathbf{x}_{\mathrm{R}\overline{j}}(l-1)$ and $\mathbf{v}_{\mathrm{R}j}(l)$, respectively. Then, in order to make the proposed selection method work for the complex-valued constellation schemes, we need to find the real-valued equivalent form of \eqref{eq09} for each symbol. To this end, let $\tilde{\mathbf{y}}^{(m)}_{\mathrm{R}j}(l)\in\mathcal{R}^{2\times1}$, $\tilde{\mathbf{x}}^{(m)}_{\mathrm{S}}(l)\in\mathcal{R}^{2\times1}$, and $\tilde{\mathbf{z}}^{(m)}_{\mathrm{R}j}(l)\in\mathcal{R}^{2\times1}$ denote real vectors obtained from $y^{(m)}_{\mathrm{R}j}(l)$, $x^{(m)}_{\mathrm{S}}(l)$, and $z^{(m)}_{\mathrm{R}j}(l)$, respectively, as
\begin{equation}\label{eq10}
\tilde{\mathbf{y}}^{(m)}_{\mathrm{R}j}(l)=[\mathcal{R}(y^{(m)}_{\mathrm{R}j}(l)),\mathcal{I}(y^{(m)}_{\mathrm{R}j}(l))]^{T},
\end{equation}
\begin{equation}\label{eq11}
\tilde{\mathbf{x}}^{(m)}_{\mathrm{S}}(l)=[\mathcal{R}(x^{(m)}_{\mathrm{S}}(l)),\mathcal{I}(x^{(m)}_{\mathrm{S}}(l))]^{T},
\end{equation}
\begin{equation}\label{eq12}
\tilde{\mathbf{z}}^{(m)}_{\mathrm{R}j}(l)=[\mathcal{R}(z^{(m)}_{\mathrm{R}j}(l)),\mathcal{I}(z^{(m)}_{\mathrm{R}j}(l))]^{T}.
\end{equation}
Additionally, let $\tilde{\mathbf{H}}_{\mathrm{S},\mathrm{R}j}(l)\in\mathcal{R}^{2\times2}$ denote real matrix obtained from $\sqrt{P}h_{\mathrm{S},\mathrm{R}j}(l)$, as
\begin{IEEEeqnarray}{ll}\label{eq13}
\tilde{\mathbf{H}}_{\mathrm{S},\mathrm{R}j}(l)=\sqrt{P}\left[\begin{array}{cc}
\mathcal{R}(h_{\mathrm{S},\mathrm{R}j}(l)) & -\mathcal{I}(h_{\mathrm{S},\mathrm{R}j}(l)) \\
\mathcal{I}(h_{\mathrm{S},\mathrm{R}j}(l)) & \mathcal{R}(h_{\mathrm{S},\mathrm{R}j}(l)) \end{array} \right]. 
\end{IEEEeqnarray}
Then, the real-valued equivalent form of \eqref{eq09} is given by
\begin{equation}\label{eq14}
\tilde{\mathbf{y}}^{(m)}_{\mathrm{R}j}(l)=\tilde{\mathbf{H}}_{\mathrm{S},\mathrm{R}j}(l)\tilde{\mathbf{x}}^{(m)}_{\mathrm{S}}(l)+\tilde{\mathbf{z}}^{(m)}_{\mathrm{R}j}(l),~\forall m.
\end{equation}
It is worth noting that, for simplicity of relays' structure and consistency with the work in \cite{Wicaksana2011}, in this paper the decoding process at the relay nodes will treat the inter-relay interference as noise. 

After obtaining the received signal as \eqref{eq14}, the relay nodes perform soft demodulation and hard-decision based decoding and then feeds the decoded bits into the same encoder and modulator as the source node to reconstruct the real-valued transmission symbol vectors $\hat{\mathbf{x}}^{(m)}_{\mathrm{S}}(l)\in\mathcal{R}^{2\times1},\forall m$.\footnote{In this paper, the repetition coded relaying is assumed for sake of simplicity, where S and R1/R2 use the same encoders and modulation scheme.} Then, each reconstructed symbol vector will be processed to decide its correctness based on the square deviation principle, which is 
\begin{equation}\label{eq15}
\Delta_m(l)=\|\mathbf{W}_{\mathrm{R}j}(l)\tilde{\mathbf{y}}^{(m)}_{\mathrm{R}j}(l)-\hat{\mathbf{x}}^{(m)}_{\mathrm{S}}(l)\|^{2},~\forall m,
\end{equation}
where $\mathbf{W}_{\mathrm{R}j}(l)\in\mathcal{R}^{2\times2}$ is the weighted matrix, which can be formulated as \cite{Kay1993}
\begin{equation}\label{eq16}
\mathbf{W}_{\mathrm{R}j}(l)=\sigma^{2}_{x}\tilde{\mathbf{H}}^{T}_{\mathrm{S},\mathrm{R}j}(l)[\sigma^{2}_{x}\tilde{\mathbf{H}}_{\mathrm{S},\mathrm{R}j}(l)\tilde{\mathbf{H}}^{T}_{\mathrm{S},\mathrm{R}j}(l)+\sigma^{2}_{z}\mathbf{I}]^{-1}.
\end{equation}
In \eqref{eq16}, $\sigma^{2}_{x}$ is covariance of a symbol element in $\tilde{\mathbf{x}}^{(m)}_{\mathrm{S}}(l)$, and $\sigma^{2}_{z}\mathbf{I}$ is covariance matrix of $\tilde{\mathbf{z}}^{(m)}_{\mathrm{R}j}(l)$, where $\mathbf{I}$ in this case is a $2\times2$ matrix. To make a hard decision whether the $m^{\mathrm{th}}$ symbol in the $l^{\mathrm{th}}$ frame is correctly decoded or not, define a utility function
\begin{IEEEeqnarray}{ll}\label{eq17}
\mathrm{sgn}\left(\varepsilon_{m}(l)\right)\triangleq\left\{
\begin{array}{l}
1,~~~~\varepsilon_{m}(l)\geq\Delta_m(l),\\
0,~~~~\mathrm{otherwise},\\
\end{array}
\right.
\end{IEEEeqnarray}
where $\varepsilon_{m}(l)$ is the square deviation error threshold between $\mathbf{W}_{\mathrm{R}j}(l)\tilde{\mathbf{y}}^{(m)}_{\mathrm{R}j}(l)$ and $\hat{\mathbf{x}}^{(m)}_{\mathrm{S}}(l)$, which is generally under the control of the system designer and should be chosen based on the prescribed modulation size. From \eqref{eq17} we can see, if $\mathrm{sgn}\left(\varepsilon_{m}(l)\right)=1$, the $m^{\mathrm{th}}$ symbols in the $l^{\mathrm{th}}$ frame is assumed to be correctly decoded and can be forwarded to the destination node, otherwise, the symbol should be transmitted with zero energy as it is determined as an erroneously decoded symbol.

With the above described symbol-level selection process, the relay node is able to formulate its actual transmission frame $\mathbf{x}_{\mathrm{R}j}(l)\in\mathcal{C}^{M}$ for the next time slot use. Specifically, let $\tilde{\mathbf{x}}^{(m)}_{\mathrm{R}j}(l)\in\mathcal{R}^{2\times1}$ be the real vector obtained from $x^{(m)}_{\mathrm{R}j}(l)$ similar to \eqref{eq11}, and $x^{(m)}_{\mathrm{R}j}(l)$ is the $m^{\mathrm{th}}$ element in $\mathbf{x}_{\mathrm{R}j}(l)$. Then, the real-valued transmission symbols in the $l^{\mathrm{th}}$ frame can be selected by
\begin{equation}\label{eq18}
\tilde{\mathbf{x}}^{(m)}_{\mathrm{R}j}(l)=\mathrm{sgn}\left(\varepsilon_{m}(l)\right)\cdot\hat{\mathbf{x}}^{(m)}_{\mathrm{S}}(l),~\forall m.
\end{equation}
After that, the real-valued $\tilde{\mathbf{x}}^{(m)}_{\mathrm{R}j}(l),\forall m,$ need to be converted back to the complex-valued $x^{(m)}_{\mathrm{R}j}(l),\forall m$. Then, the actual transmission frame $\mathbf{x}_{\mathrm{R}j}(l)$ is obtained.

After receiving the data frames during two successive time slots, the destination node will implement the modified MAP receiver to mitigate inter-frame interference generated from both source and relay nodes, and identify the position of discarded symbols. In detail, in comparison to the standard MAP receiver, the modified MAP receiver allows the hypothesis-detection include the event that the symbol sent from the relay nodes could be with zero energy. Then, after splitting the two successive frames, the position of the discarded symbols from the relay nodes can be automatically identified. Subsequently, the two versions of one data frame can be combined for the turbo-like decoding process.

\section{Diversity Multiplexing Trade-off Analysis}
To exploit the DMT performance of our proposed scheme, the probability of correctly predicted/forwarded symbols per frame at the relay nodes need to be calculated first. Such probability can be derived with the help of the statistical property of estimation error in \eqref{eq15}. Specifically, we assume all the channel links are complex Gaussian distributed, and from \eqref{eq14} we can see, $\tilde{\mathbf{H}}_{\mathrm{S},\mathrm{R}j}(l)$ is fixed for the $l^{\mathrm{th}}$ time slot, and $\tilde{\mathbf{z}}^{(m)}_{\mathrm{R}j}(l)$ follows normal distribution, i.e., $\mathcal{N}(0,\sigma^{2}_{z}\mathbf{I})$. Thus, according to \cite{Kay1993}, the estimation error vector (i.e. $\mathbf{W}_{\mathrm{R}j}(l)\tilde{\mathbf{y}}^{(m)}_{\mathrm{R}j}(l)-\hat{\mathbf{x}}^{(m)}_{\mathrm{S}}(l)$) follows normal distribution with zero mean and covariance matrix as 
\begin{IEEEeqnarray}{ll}\label{eq19}
\mathbf{C}_{e}\triangleq\nonumber\\
\sigma^{2}_{x}\mathbf{I}-\sigma^{4}_{x}\tilde{\mathbf{H}}^{T}_{\mathrm{S},\mathrm{R}j}(l)[\sigma^{2}_{x}\tilde{\mathbf{H}}_{\mathrm{S},\mathrm{R}j}(l)\tilde{\mathbf{H}}^{T}_{\mathrm{S},\mathrm{R}j}(l)+\sigma^{2}_{z}\mathbf{I}]^{-1}\tilde{\mathbf{H}}_{\mathrm{S},\mathrm{R}j}(l),\nonumber\\
\end{IEEEeqnarray}
where, in \eqref{eq19}, $\tilde{\mathbf{H}}_{\mathrm{S},\mathrm{R}}(l)\tilde{\mathbf{H}}^{T}_{\mathrm{S},\mathrm{R}}(l)=P_{\mathrm{S}}\sigma^{2}_{h_{\mathrm{S,R}}}\mathbf{I}$ is a scalar matrix, so that \eqref{eq19} can be simplified as 
\begin{IEEEeqnarray}{ll}\label{eq19a}
\mathbf{C}_{e}=\underbrace{\left(\sigma^{2}_{x}-\frac{P_{\mathrm{S}}\sigma^{4}_{x}\sigma^{2}_{h_{\mathrm{S,R}}}}{P_{\mathrm{S}}\sigma^{2}_{x}\sigma^{2}_{h_{\mathrm{S,R}}}+\sigma^{2}_{z}}\right)}_{\triangleq\sigma^{2}_{C_e}}\mathbf{I},
\end{IEEEeqnarray}
which is also a scalar matrix with the scalar value being defined as $\sigma^{2}_{C_e}$. Thus, the square deviation $\Delta_m(l)$ follows the chi-squared distribution with two degrees of freedom. In this case, the probability of the $m^{\mathrm{th}}$ symbol in the $l^{\mathrm{th}}$ frame that is selected to be forwarded to the destination node is given by
\begin{eqnarray}\label{eq20}
\mathcal{P}_{m}(l)&\triangleq&\mathrm{Pr}(\Delta_m(l)\leq\varepsilon_{m}(l)),\nonumber\\
&=&\int^{\frac{\varepsilon_{m}(l)}{\sigma^{2}_{C_e}}}_{0}f_{\Delta}(x;2)dx,\nonumber\\
&=&1-e^{-\frac{\varepsilon_{m}(l)}{2\sigma^{2}_{C_e}}},
\end{eqnarray}
where $f_{\Delta}(x;2)$ in \eqref{eq20} is the probability density function (p.d.f.) of $\Delta_m(l)$. Due to the statistical independence, the average probability of correctly predicted/forwarded symbols per frame at the virtual FD relay nodes is given by 
\begin{eqnarray}\label{eq20a}
\mathcal{P}_{\mathrm{C}}\triangleq\frac{1}{LM}\sum^{L}_{l=1}\sum^{M}_{m=1}\mathcal{P}_{m}(l).
\end{eqnarray}
It is worth noting that, if a specific symbol is discard at one relay in the current time slot, the decision made for the corresponding symbol at the other relay in the next time slot will not be affected by the inter-relay interference. In this case, when we calculate the selection probability for the corresponding symbol, $\sigma^{2}_{z}$ in \eqref{eq19a} will not include the variance of inter-relay interference.

\begin{figure}[t] 
\begin{center}
\epsfig{figure=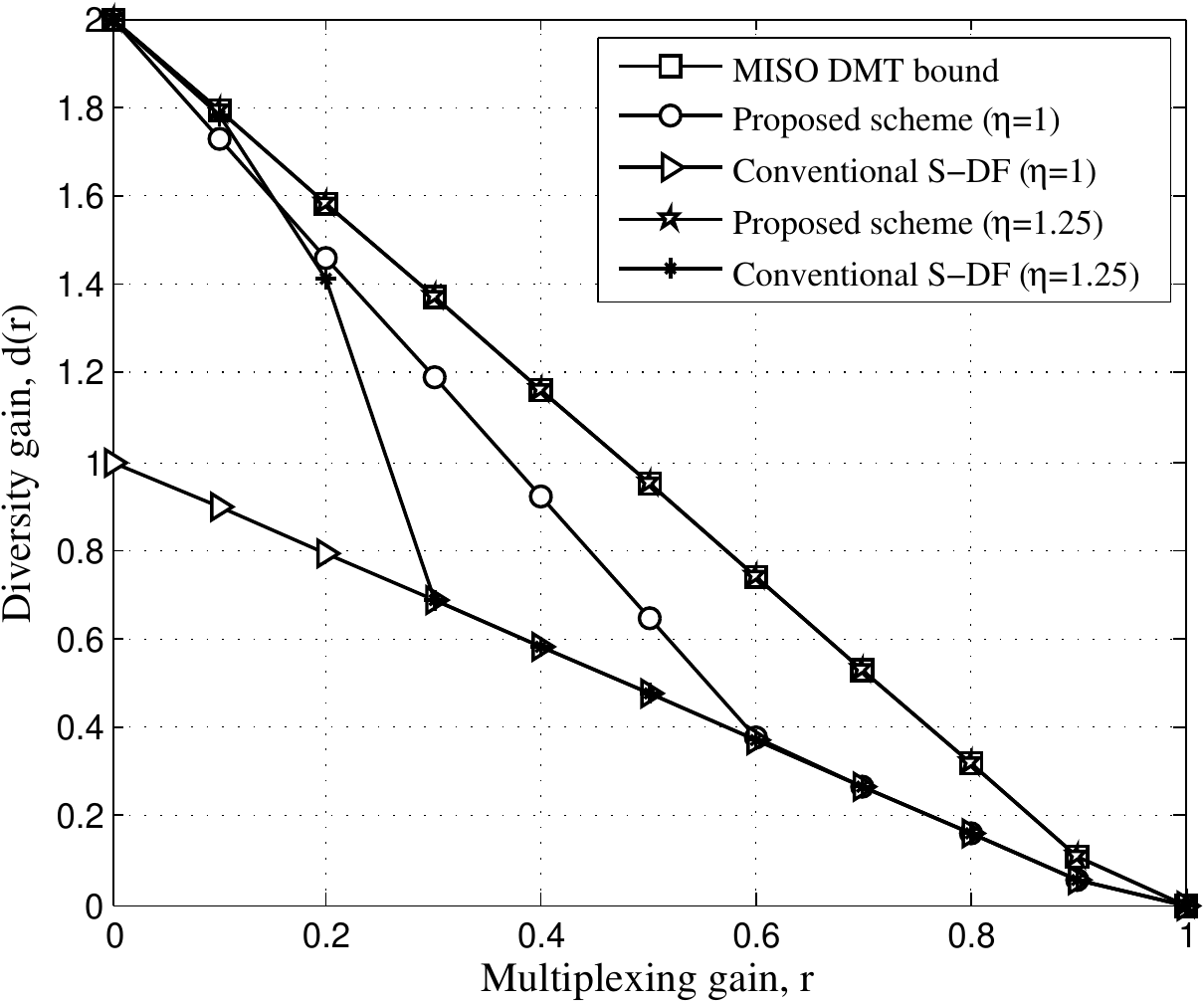,scale=0.66,angle=0}
\end{center}
\caption{Diversity-multiplexing trade-off for different virtual FD relaying schemes, where $L=20$ and $\varepsilon_m(l)=0.5,\forall m, l$.}\label{F2}
\end{figure}

After obtained the average probability of correctly selected symbols per frame at the relay nodes, the DMT expression of our proposed scheme can be derived following the derivation procedure in the Appendix of \cite{Fan2007} and can be expressed as
\begin{equation}\label{eq23}
d(r)=\left(1-\frac{L+1}{L}r\right)^{+}+\left(1-\frac{L+1}{L\mathcal{P}_{\mathrm{C}}}r\right)^{+},
\end{equation}
where $d$ represents spatial diversity gain and $r$ represents spatial multiplexing gain; $(a)^{+}$ denotes $\max\{0,a\}$. Let's define the average SNR of S-R link over the average SNR of R-R link as $\eta$, where SNR is defined as the transmission power normalized by noise power ratio. Fig.~\ref{F2} compares the DMT performances of different schemes with different $\eta$ values, where \textit{MISO DMT bound} is the optimal DMT performance derived in \cite{Fan2007}, and \textit{Conventional S-DF} is the DMT performance for CRC based S-DF relaying in \cite{Wicaksana2011}.
As shown in Fig.~\ref{F2}, \textit{MISO DMT bound} scheme gives the best DMT performance because the perfect decoding at the relays is assumed. For the case $\eta=1$, where the average SNR of S-R link is the same as the average SNR of R-R link, our proposed scheme outperforms \textit{Conventional S-DF} scheme when the multiplexing gain $r$ is less than 0.6. This is because, in this case, the relay nodes for \textit{Conventional S-DF} scheme always discard the received frames due to the decoding errors. On the other hand, the relay nodes for our proposed scheme can base on the symbol-level selection method to predict and forward the correctly decoded symbols per frame to the destination node to enjoy partial spacial diversity gain. For the case $\eta=1.25$, where the average SNR of S-R link is 1.25 times better than the average SNR of R-R link, the DMT performance of \textit{Conventional S-DF} scheme has some improvements when $r$ is less than 0.3. In this case, the DMT performance of our proposed scheme is almost the same as the one of \textit{MISO DMT bound} scheme. This is because, with $\eta=1.25$, the effects of inter-relay interference for our proposed scheme can be ignored especially when the average SNR of diffident links goes to infinity.



\section{Simulation Results}
Computer simulations are conducted to evaluate our proposed symbol-level selective transmission scheme in terms of BER performances. We assume that all channels are generated as independent Rayleigh fading, which remained static over each time slot. $L=20$ frames are transmitted via $L+1$ time slots, and each frame is with $M=512$ information bits. The quadrature phase-shift keying (QPSK) modulation and turbo-like channel coding are implemented in the system. Specifically, the 1/2 rate serial concatenated convolutional code is used at the source and relay nodes, where the first encoder is the non-recursive non-systematic convolutional code with a generator polynomial $G=([3,2])_{8}$, and the second encoder is the doped-accumulator with a doping rate equalling 2. Simulation results are computed on an average over 1000 independent channel realizations.  

There are three baselines for comparison: 1) \textit{Perfect decoding at relays}: this is served as the performance bound, where we assume that the relay nodes can always perfectly decode the source node's messages; 2) \textit{CRC based S-DF}: the relay nodes only forward if they can perfectly decode the source node's messages; 3) \textit{Threshold based S-DF}: the relay nodes only forward if the probability of decoding errors in a frame is less than 10\%. All the schemes implement our modified MAP detector at the destination node to mitigate inter-frame interference. In addition, given the distance between the source and the destination nodes as $d_{\mathrm{S,D}}=d$, we have relay location as $d_{\mathrm{S},\mathrm{R1}}=d_{\mathrm{S,R2}}=\frac{1}{2}d$ and $d_{\mathrm{R1,D}}=d_{\mathrm{R2,D}}=\frac{3}{4}d$. Then, following the work in \cite{Anwar2012}, the SNR relationship in dB among different links can be approximated by $\rho_{\mathrm{S,R1}}=\rho_{\mathrm{S,R2}}=\rho_{\mathrm{S,D}}+10.6~\mathrm{dB}$ and $\rho_{\mathrm{R1,D}}=\rho_{\mathrm{R2,D}}=\rho_{\mathrm{S,D}}+4.4~\mathrm{dB}$. This configuration illustrates the relay nodes are close to the source node, which is suitable for DF relaying.

\begin{figure}[t] 
\begin{center}
\epsfig{figure=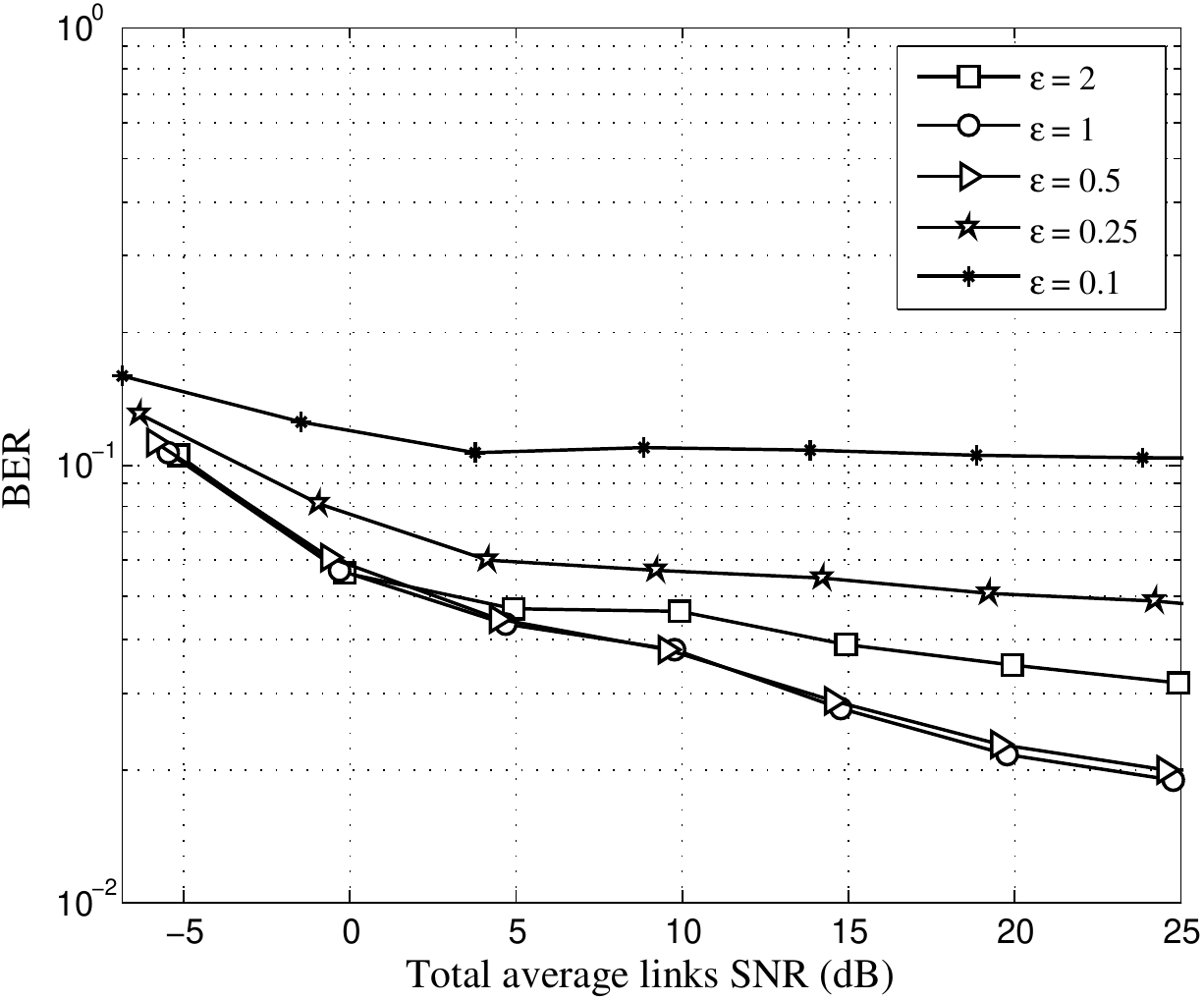,scale=0.66,angle=0}
\end{center}
\caption{BER versus the total average links SNR for the proposed scheme with different $\varepsilon$ values, where the variance of inter-relay link was set up to 1.}\label{F3}
\end{figure}

\textit{Experiment 1:} As we mentioned in Section III, $\varepsilon_{m}(l),\forall m,l,$ affect symbol prediction accuracy and should be carefully designed. In this experiment, we will base on computer simulations to pick up suitable $\varepsilon_{m}(l),\forall m,l$. Due to statistical identity, we have $\varepsilon_{m}(l)=\varepsilon,\forall m,l$. Fig.~\ref{F3} gives BER performances of our proposed scheme versus the total average links SNR for different $\varepsilon$ values.
Here, we choose the variance of inter-relay channel link equalling to one as an example. For other inter-relay channel variance configurations, the same performance trend can be observed. The total average link SNR denotes the total transmission power consumed at both transmitters to noise power ratio divided by two. As shown in Fig.~\ref{F3}, the case where $\varepsilon=1$ give the best BER performance by comparing with the other cases, which means giving a too large or a too small value of $\varepsilon$ are not suitable for predicting the correctly decoded symbols at the relay nodes. Here, if $\varepsilon$ is too large, although more symbols are selected, the symbols can be wrongly decoded; if $\varepsilon$ is too small, even some symbols are correctly decoded, the threshold will stop the symbols being selected. For the next experiment, $\varepsilon=1$ will be used for the BER performance evaluation.

\begin{figure}[t] 
\begin{center}
\epsfig{figure=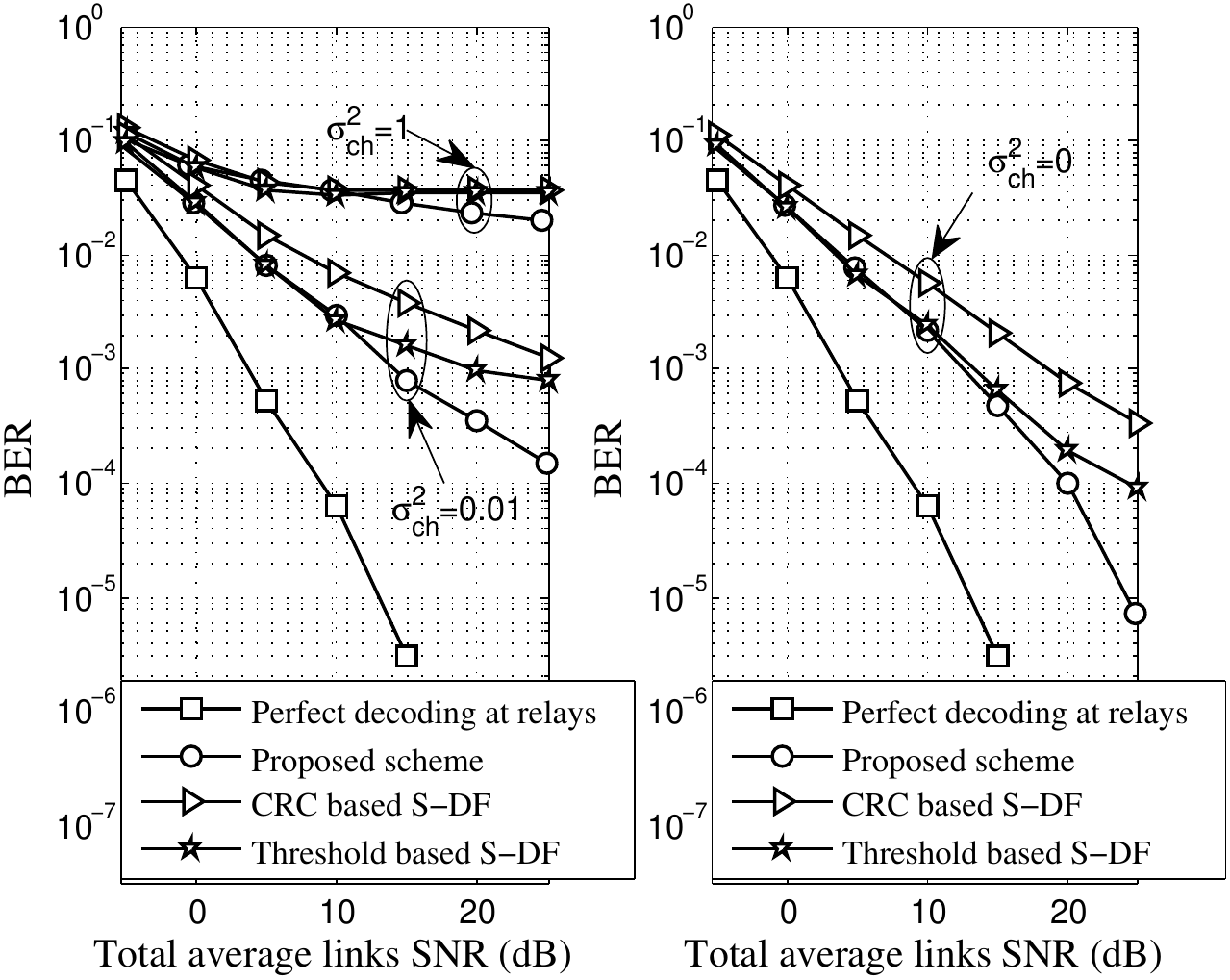,scale=0.66,angle=0}
\end{center}
\caption{BER versus the total average links SNR for different kinds of relaying schemes, where the variance of inter-relay link (i.e. $\sigma^2_{\mathrm{ch}}$) was set up to 1, 0.01, and 0, respectively.}\label{F4}
\end{figure}

\textit{Experiment 2:} The objective of this experiment is to examine BER performances of our proposed symbol-level selective scheme with different levels of inter-relay interference. Fig.~\ref{F4} gives BER performances versus the total average links SNR for different kinds of relaying schemes with different channel variances of inter-relay link.
In this case, $\sigma^2_{\mathrm{ch}}$ denotes the channel variance of the inter-relay link. If $\sigma^2_{\mathrm{ch}}=1$, S-R link and R-R link have the same quality; if $\sigma^2_{\mathrm{ch}}=0.01$, the inter-relay interference does exist but not as strong as S-R link; if $\sigma^2_{\mathrm{ch}}=0$, there is no inter-relay interference. To clarify the comparison of different schemes with different cases, we split one figure to two sub-figures.

As shown in Fig.~\ref{F4}, the scheme, that both relay nodes can always perfectly decode the source node messages, gives the best BER performance and enjoys the full diversity gain. Our proposed scheme gives the second best BER performance by comparing with the other existing schemes especially for high SNR range. Such performance trend is in line with the DMT performances discussed in Section IV. Specifically, for the case where $\sigma^2_{\mathrm{ch}}=1$, our proposed scheme outperforms both \textit{CRC based S-DF} scheme and \textit{Threshold based S-DF} scheme when the total average links SNR is larger than 12 dB. For the case where $\sigma^2_{\mathrm{ch}}=0.01$, our proposed scheme and \textit{Threshold based S-DF} scheme outperform \textit{CRC based S-DF} scheme through the entire displayed SNR range. Apart from that, our proposed scheme outperforms \textit{Threshold based S-DF} scheme when the SNR is larger than 10 dB. This is because that our proposed scheme can always select the correctly decoded symbols and forward them to enjoy partial diversity gain. However, \textit{CRC based S-DF} scheme stops forwarding even there is only a single erroneously decoded symbol in a frame. On the other hand, \textit{Threshold based S-DF} scheme can tolerate certain amount of erroneously decoded symbols to enjoy increased spacial diversity gain by comparing with \textit{CRC based S-DF} scheme. For the case where $\sigma^2_{\mathrm{ch}}=0$, the same performance trend can be observed as the case where $\sigma^2_{\mathrm{ch}}=0.01$. As shown in the second sub-figure, because the inter-relay interference in this case is equal to zero, our proposed scheme can offer the same diversity gain as \textit{Perfect decoding at relays} scheme.



\section{Conclusion}
In this paper, a square deviation based symbol-level selective transmission scheme for virtual FD relaying has been proposed. This scheme predicts the correctly decoded symbols at the relay nodes and then forwarded them to the destination node to improve the system spatial diversity gain. Apart from that, the DMT performance of our proposed scheme has been theoretically analysed, and numerical results show that our proposed scheme outperforms the conventional CRC based S-DF relaying scheme in terms of DMT. Moreover, BER performances for different schemes have also been simulated to confirm the DMT results.



\section*{Acknowledgment}
This work was supported by the Engineering and Physical Science Research Council (EPSRC) through the Scalable Full Duplex Dense Wireless Networks (SENSE) grant EP/P003486/1.
\

\bibliography{mybib}

\begin{thebibliography}{10}
\providecommand{\url}[1]{#1}
\csname url@samestyle\endcsname
\providecommand{\newblock}{\relax}
\providecommand{\bibinfo}[2]{#2}
\providecommand{\BIBentrySTDinterwordspacing}{\spaceskip=0pt\relax}
\providecommand{\BIBentryALTinterwordstretchfactor}{4}
\providecommand{\BIBentryALTinterwordspacing}{\spaceskip=\fontdimen2\font plus
\BIBentryALTinterwordstretchfactor\fontdimen3\font minus
  \fontdimen4\font\relax}
\providecommand{\BIBforeignlanguage}[2]{{%
\expandafter\ifx\csname l@#1\endcsname\relax
\typeout{** WARNING: IEEEtran.bst: No hyphenation pattern has been}%
\typeout{** loaded for the language `#1'. Using the pattern for}%
\typeout{** the default language instead.}%
\else
\language=\csname l@#1\endcsname
\fi
#2}}
\providecommand{\BIBdecl}{\relax}
\BIBdecl

\bibitem{Laneman2004}
J.~N. Laneman, D.~N.~C. Tse, and G.~W. Wornell, ``Cooperative diversity in
  wireless networks: Efficient protocols and outage behavior,'' \emph{IEEE
  Trans. Inf. Theory}, vol.~50, no.~12, pp. 3062--3080, Dec. 2004.

\bibitem{Chen2006}
D.~Chen and J.~N. Laneman, ``Modulation and demodulation for cooperative
  diversity in wireless systems,'' \emph{IEEE Trans. Wireless Commun.}, vol.~5,
  no.~7, pp. 1785--1794, Jul. 2006.

\bibitem{Lai2006}
L.~Lai, K.~Liu, and H.~E. Gamal, ``The three-node wireless network: Achievable
  rates and cooperation strategies,'' \emph{IEEE Trans. Inf. Theory}, vol.~52,
  no.~3, pp. 805--828, Mar. 2006.

\bibitem{Oechtering2004}
T.~J. Oechtering and A.~Sezgin, ``A new cooperative transmission scheme using
  the space-time delay code,'' in \emph{Proc. ITG Wksp. Smart Antennas}, Mar.
  2004, pp. 41--48.

\bibitem{Ribeiro2004}
A.~Ribeiro, X.~Cai, and G.~B. Giannakis, ``Opportunistic multipath for
  bandwidth-efficient cooperative networking,'' in \emph{Proc. IEEE Int. Conf.
  Acoustics, Speech and Signal Processing}, May 2004, pp. 549--552.

\bibitem{Ju2009}
H.~Ju, E.~Oh, and D.~Hong, ``Improving efficiency of resource usage in two-hop
  full duplex relay systems based on resource sharing and interference
  cancellation,'' \emph{IEEE Trans. Wireless Commun.}, vol.~8, no.~8, pp.
  3933--3938, Aug. 2009.

\bibitem{Duarte2010}
M.~Duarte and A.~Sabharwal, ``Full-duplex wireless communications using
  off-the-shelf radios: feasibility and first results,'' in \emph{Proc. 44th
  Asilomar Conf. Signals Syst. Comput.}, Nov. 2010, pp. 1558--1562.

\bibitem{Naslcheraghi2017}
M.~Naslcheraghi, S.~A. Ghorashi, and M.~Shikh-Bahaei, ``{FD} device-to-device
  communication for wireless video distribution,'' \emph{IET Commun.}, vol.~11,
  no.~7, pp. 1074--1081, May 2017.

\bibitem{Fan2007}
Y.~Fan, C.~Wang, J.~S. Thompson, and H.~V. Poor, ``Recovering multmultiple loss
  through successive relaying using repetition coding,'' \emph{IEEE Trans.
  Wireless Commun.}, vol.~6, no.~12, pp. 4484--4493, Dec. 2007.

\bibitem{Kobravi2007}
A.~Kobravi and M.~Shikh-Bahaei, ``Cross-layer adaptive {ARQ} and modulation
  tradeoffs,'' in \emph{Proc. IEEE Int. Symp. Personal, Indoor and Mobile Radio
  Communications}, Sept. 2007, pp. 1--5.

\bibitem{Wang2009}
C.~Wang, Y.~Fan, J.~S. Thompson, and H.~V. Poor, ``A comprehensive study of
  repetition-coded protocol in multi-user multi-relay networks,'' \emph{IEEE
  Trans. Wireless Commun.}, vol.~8, no.~8, pp. 4329--4339, Aug. 2009.

\bibitem{Bobarshad2009}
H.~Bobarshad and M.~Shikh-Bahaei, ``M/m/1 queuing model for adaptive
  cross-layer error protection in wlans,'' in \emph{Proc. IEEE Wireless Commun.
  and Networking Conf.}, Apr. 2009, pp. 1--6.

\bibitem{Shikh-Bahaei2004}
M.~Shikh-Bahaei, M.~Mouna-Kingue, and G.~C. Nokia, ``Joint optimisation of
  outer-loop power control and rate adaptation over fading channels,'' in
  \emph{Proc. IEEE Veh. Tech. Conf..}, Sept. 2004, pp. 2205--2209.

\bibitem{Tian2011}
F.~Tian, W.~Zhang, W.~K. Ma, P.~C. Ching, and H.~V. Poor, ``An effective
  distributed space-time code for two-path successive relay network,''
  \emph{IEEE Trans. Commun.}, vol.~59, no.~8, pp. 2254--2263, Aug. 2011.

\bibitem{Shadmand2010}
A.~Shadmand and M.~Shikh-Bahaei, ``Multi-user time-frequency downlink
  scheduling and resource allocation for {LTE} cellular systems,'' in
  \emph{Proc. IEEE Wireless Commun. and Networking Conf.}, Apr. 2010, pp. 1--6.

\bibitem{Hu2012}
Y.~Hu, K.~H. Li, and K.~C. Teh, ``An efficient successive relaying protocol for
  multiple-relay cooperative networks,'' \emph{IEEE Trans. Wireless Commun.},
  vol.~11, no.~5, pp. 1892--1899, May 2012.

\bibitem{Shadmand2009}
A.~Shadmand and M.~Shikh-Bahaei, ``{TCP} dynamics and adaptive {MAC}
  retry-limit aware link-layer adaptation over ieee 802.11 wlan,'' in
  \emph{Proc. Seventh Annual Commun. Networks and Services Research Conf.}, May
  2009, pp. 193--200.

\bibitem{Zarringhalam2009}
F.~Zarringhalam, B.~Seyfe, M.~Shikh-Bahaei, G.~Charbit, and H.~Aghvami,
  ``Jointly optimized rate and outer loop power control with single- and
  multi-user detection,'' \emph{IEEE Trans. Wireless Commun.}, vol.~8, no.~1,
  pp. 186--195, Jan. 2009.

\bibitem{Olfat2005}
A.~Olfat and M.~Shikh-Bahaei, ``Optimum power and rate adaptation with
  imperfect channel estimation for mqam in rayleigh flat fading channel,'' in
  \emph{Proc. IEEE Veh. Tech. Conf.}, Sept. 2005, pp. 2468--2471.

\bibitem{Shojaeifard2011}
A.~Shojaeifard, F.~Zarringhalam, and M.~Shikh-Bahaei, ``Joint physical layer
  and data link layer optimization of cdma-based networks,'' \emph{IEEE Trans.
  Wireless Commun.}, vol.~10, no.~10, pp. 3278--3287, Oct. 2011.

\bibitem{Sun2011}
L.~Sun, T.~Zhang, and H.~Niu, ``Inter-relay interference in two-path digital
  relaying systems: Detrimental or beneficial?'' \emph{IEEE Trans. Wireless
  Commun.}, vol.~10, no.~8, pp. 2468--2473, Aug. 2011.

\bibitem{Basar2014}
E.~Basar, U.~Aygolu, E.~Panayirci, and H.~V. Poor, ``A reliable successive
  relaying protocol,'' \emph{IEEE Trans. Commun.}, vol.~62, no.~5, pp.
  1431--1443, May 2014.

\bibitem{Wicaksana2011}
H.~Wicaksana, S.~H. Ting, Y.~L. Guan, and X.-G. Xia, ``Decode-and-forward
  two-path half-duplex relaying: Diversity-multiplexing tradeoff analysis,''
  \emph{IEEE Trans. Commun.}, vol.~59, no.~7, pp. 1985--1994, Jul. 2011.

\bibitem{Kay1993}
S.~M. Kay, \emph{Fundamentals of Statistical Signal Processing: Estimation
  Theory}.\hskip 1em plus 0.5em minus 0.4em\relax New Jersey: Prentice-Hall,
  Inc., 1993, vol.~1.

\bibitem{Al-Habian2011}
G.~Al-Habian, A.~Ghrayeb, M.~Hasna, and A.~Abu-Dayya, ``Threshold-based
  relaying in coded cooperative networks,'' \emph{IEEE Trans. Veh. Technol.},
  vol.~60, pp. 123--135, Jan. 2011.

\bibitem{Kwon2010}
T.~Kwon, S.~Lim, S.~Choi, and D.~Hong, ``Optimal duplex mode for df relay in
  terms of the outage probability,'' \emph{IEEE Trans. Veh. Technol.}, vol.~59,
  no.~7, pp. 3628--3633, Sept. 2010.

\bibitem{Yi2015}
N.~Yi, Y.~Ma, J.~Hou, and R.~Tafazolli, ``Symbol-level selective decode-forward
  relaying for uncoordinated dense wireless networks,'' in \emph{Proc. IEEE
  Int. Wksp. Computer Aided Modelling and Design of Communication Links and
  Networks}, Sept. 2015, pp. 283--287.

\bibitem{Anwar2012}
K.~Anwar and T.~Matsumoto, ``Accumulator-assisted distributed turbo codes for
  relay systems exploiting source-relay correlation,'' \emph{IEEE Commun.
  Lett.}, vol.~16, no.~7, pp. 1114--1117, Jul. 2012.

\end{thebibliography}
\bibliographystyle{IEEEtran}
\end{document}